\RequirePackage[2014/01/01]{latexrelease}
\documentclass[sigconf,usenames,dvipsnames]{acmart}

\usepackage[T1]{fontenc}
\usepackage{booktabs} 
\usepackage{tabularx}
\newcolumntype{L}[1]{>{\hsize=#1\hsize\raggedright\arraybackslash}X}%
\newcolumntype{R}[1]{>{\hsize=#1\hsize\raggedleft\arraybackslash}X}%
\newcolumntype{C}[1]{>{\hsize=#1\hsize\centering\arraybackslash}X}%

\makeatletter
\def\subsubsection{\@startsection{subsubsection}{3}%
  \z@{.5\linespacing\@plus.7\linespacing}{.1\linespacing}%
  {\normalfont\itshape}}
\makeatother

\usepackage{url}
\usepackage{algorithm}
\usepackage{algorithmic}
\usepackage{graphicx}
\usepackage{subcaption}
\usepackage{caption}
\usepackage{soul}
\usepackage{color}
\usepackage{xcolor}
\usepackage{multirow}
\usepackage{fixltx2e}  
\usepackage{csquotes}
\usepackage[nodayofweek,level]{datetime}
\usepackage{tikz}
\usetikzlibrary{arrows}
\usepackage{balance}
\usepackage{changebar}

\usepackage{amsmath}

\newcommand{\hlc}[2][yellow]{{\sethlcolor{#1} \hl{#2}}}

\newcommand{\mname}{TOP-ID} 

\definecolor{pink0}{HTML}{f9e7f0}
\definecolor{pink1}{HTML}{f6cee2}
\definecolor{pink2}{HTML}{ffa2d0}
\definecolor{pink3}{HTML}{ff4aa4}
\definecolor{pink4}{HTML}{e30372}

\DeclareMathOperator*{\argmax}{arg\,max}

\setcopyright{rightsretained}

\copyrightyear{2018}
\acmYear{2018}
\setcopyright{acmlicensed}
\acmPrice{15.00}
\acmDOI{10.475/123_4}
\acmISBN{123-4567-24-567/08/06}

\begin{document}

\title{Towards Open Intent Discovery for Conversational Text}
\author{Nikhita Vedula}
\affiliation{%
  \institution{Ohio State University}
}
\email{vedula.5@osu.edu}

\author{Nedim Lipka}
\affiliation{%
  \institution{Adobe Research}
}
\email{lipka@adobe.com}

\author{Pranav Maneriker}
\affiliation{%
  \institution{Ohio State University}
}
\email{maneriker.1@osu.edu}

\author{Srinivasan Parthasarathy}
\affiliation{%
  \institution{Ohio State University}
}
\email{srini@cse.ohio-state.edu}

\begin{abstract}

Detecting and identifying user intent from text, both written and spoken, plays an important role in modelling and understand dialogs. 
Existing research for intent discovery model it as a classification task with a predefined set of known categories. 
To generailze beyond these preexisting classes, we define a new task of \textit{open intent discovery}. 
We investigate how intent can be generalized to those not seen during training. 
To this end, we propose a two-stage approach to this task - predicting whether an utterance contains an intent, and then tagging the intent in the input utterance. 
Our model consists of a bidirectional LSTM with a CRF on top to capture contextual semantics, subject to some constraints. 
Self-attention is used to learn long distance dependencies. Further, we adapt an adversarial training approach to improve robustness and perforamce across domains. 
We also present a dataset of 25k real-life utterances that have been labelled via crowd sourcing. 
Our experiments across different domains and real-world datasets show the effectiveness of our approach, with less than 100 annotated examples needed per unique domain to recognize diverse intents. 
The approach outperforms state-of-the-art baselines by 5-15\% F1 score points.

\keywords{}

\end{abstract}

\maketitle

\section{Introduction}
\label{sec:introduction}

Recent advances in the efficacy and accuracy of natural language understanding (NLU) and speech recognition technologies have triggered the advent of a wealth of conversational agents such as Apple's Siri, Microsoft's Cortana and Amazon's Alexa. In order to effectively and intelligently interact with people and answer their diverse questions, such agents need to parse and interpret human language utterances, especially people's intentions, and respond accordingly. The problem of recognizing human intentions or \textit{intents} from their text or speech inputs has several downstream applications. It can help in summarizing the common or frequent user objectives and functions associated with a business or a product. It can highlight and help prioritize common bugs and issues reported to customer support or public forums, and spot action items in emails or meeting transcripts.
Progress in the field of deep learning has led to the emergence of models that can detect user intents~\cite{chen2013identifying,bhargava2013,xu2013convolutional,gupta2014identifying,wang2015mining,kim2016intent,liu2016attention,zhang2016joint,kim2017onenet,goo2018slot,coucke2018,xia2018} and identify semantically relevant entities linked with those intents (slot filling). 

Most existing research including commercial NLU engines formulate the problem of identifying user intents as a multi-class classification task. Assuming the presence of an intent in a given user utterance, such models categorize the utterance into pre-defined intent classes for which sufficient labeled data is available during the model training phase. Most techniques are unable to address inputs that belong to new or previously unseen intent categories, i.e., they work with a \textit{closed world} assumption. 
Studies further assume that an input text expresses only a single intent (e.g.~\cite{liu2016attention,kim2016intent,goo2018slot,coucke2018,xia2018}). This is very much unlike real-world scenarios where users often express multiple, distinct intentions within one dialog turn or utterance.

In our work, we propose a framework called \mname{} (\underline{T}owards \underline{OP}en \underline{I}ntent \underline{D}iscovery). It automatically discovers user intents in natural language \textit{without} prior knowledge of a comprehensive list of intent classes that the text may comprise of. In other words, \mname{} is not restricted to a pre-defined set of intent categories. It can recognize instances of intent types that it has never seen before. This is a much more challenging problem than the multi-class classification problem that prior literature generally formulates the intent detection task as.
We therefore name this novel task as \textit{Open Intent Discovery}. It focuses on identifying and extracting user intentions from text utterances \textit{explicitly} containing them in their content. It does not infer or deduce the intent if it is implicitly stated in the text. 
To illustrate, the text \textit{``Please make a 10:30 sharp appointment for a haircut"} contains a single intent of making a haircut appointment; whereas the text \textit{``I would like to reserve a seat and also if possible, request a special meal on my flight"} contains multiple intents -- a seat reservation and a meal request.
Contrarily, the sentence \textit{``Anyone knows the battery life of iPhone?"} merely requests information on a particular topic and does not seem to contain any tangible intent action, such as that of buying an iPhone. We do not consider such ambiguous or questionable utterances.

Recent work by Xia et al~\cite{xia2018} has a similar objective as ours, i.e., recognizing intents outside of the labeled training data. They treat this as a zero-shot classification problem.
However, their method can only handle the basic case of an input utterance containing a single intent. It also requires a list of new or unseen intent classes to be available at test time, and classifies the text input into one of them.  
Our work does not have these restrictions, and to the best of our knowledge, is the \textit{first} work to address the aforementioned limitations. It gives a fine-grained picture of the diverse user intents in an utterance or a collection of utterances, rather than merely grouping intents into higher level categories.

Our proposed two-stage approach, TOP-ID, aims to solve the problem of open intent discovery.
In the first stage, our method employs a softmax classifier on top of a bidirectional LSTM to determine if the text input is likely to contain an explicit, tangible intent or not. 
If it does, the second stage of \mname{} is applied to identify and extract all possible intents in a consistent and generalizable format. We model this as a \textit{sequence tagging} problem. We solve it by developing a neural network model consisting of a Conditional Random Field (CRF) on top of a bidirectional LSTM, accompanied by a multi-head self-attention mechanism.
A crucial challenge associated with developing a generic technique for open intent discovery is ensuring its effectiveness it across several task domains or fields. 
\mname{} represents all kinds of user intents extracted from the textual input in a common format, independent of their domain. We further employ adversarial training at the lower layers of our model, and pre-train it without supervision in the target domain under consideration. 
These strategies empower our model for cross-domain adaptation even in the absence of sufficient labeled training data, as we show empirically in Section~\ref{sec:evaluation}. 
Moreover, commonly used intent-labeled datasets in dialog research such as SNIPS~\cite{coucke2018} or ATIS~\cite{hemphill1990,dahl1994} largely have concise, coherent and single-sentence texts. They are not very representative of complex, real-world dialog scenarios which could be verbose and ungrammatical, with intents scattered throughout their content. 
Therefore, we develop a large dataset with 25K real-world utterances from the online question-answer forum of Stack Exchange. They span several genres and have been curated for intents by crowd workers. 

To summarize, the key contributions of our work are:

\begin{itemize}
\item We formulate and solve the novel problem of \textit{open intent discovery} in text utterances. Our proposed two-stage technique \mname{} is flexible, generalizable, and agnostic of the domain of the target text. 
\item \mname{}, can discover both previously seen as well as unseen (during training) user intents in diverse real-world scenarios. It can identify multiple user intents per utterance.
\item We curate and present a large, intent-annotated dataset of 25K text instances from real-world task domains, without any restriction on the number or types of intents possible.

\end{itemize}
\section{Related Work}
\label{sec:background}


Prior work on user intents encompasses two avenues: asynchronous, written communication (forums, blogs, tweets) and synchronous dialog. In both cases, intent detection is frequently modeled as a binary or multi-class classification problem, with each class representing the presence or absence of a specific kind of intent. 
Supervised and semi-supervised learning models based on linguistic and sentiment features have been used to model racial intent on Tumblr~\cite{agarwal2017characterizing}, and purchase intent in social posts~\cite{gupta2014identifying,wang2015mining} and in discussion forums~\cite{chen2013identifying}. 
Modeling responses for dialog agents such as Microsoft LUIS\footnote{https://www.luis.ai/}, Google Dialogflow\footnote{https://dialogflow.com/}, and Amazon Lex\footnote{https://aws.amazon.com/lex/} includes the dual tasks of domain-aware intent detection and slot filling. 
Recent approaches have drawn upon progress in CNN, RNN, and CRF based language models to improve intent detection~\cite{xu2013convolutional,mesnil2015using,kim2016intent,liu2016attention,zhang2016joint,kim2017onenet,goo2018slot}. 
Performing slot filling jointly with intent detection has improved the performance of both tasks~\cite{kim2016intent,liu2016attention,zhang2016joint}. 

In the health community, Cai et al~\cite{cai2017cnn} used hierarchical clustering to learn a taxonomy of intent classes, and applied a hybrid CNN-LSTM model to classify the intent of medical queries. We take this idea further and learn to identify arbitrary intents beyond even a predefined taxonomy. Improvements in intent detection and slot filling based on adversarial learning have also been explored~\cite{liu2017multi,kim2017adversarial,yu2018product}. 
We exploit adversarial training to generate adversarial input examples for improving the performance of our model and for domain adaptation (Section~\ref{sec:adversarial}).
Moreover, adding linguistic structure to existing models has been shown to improve their performance across a wide range of NLU tasks, such as word embedding~\cite{mrkvsic2016counter}, machine translation~\cite{chen2017improved}, named entity recognition~\cite{jochim2017named}, and semantic role labelling~\cite{he2017deep}. 
We impose linguistic constraints on the CRF layer of \mname{} to preserve the semantics of intent actions and their associated objects (Section~\ref{sec:constraint}).

\section{Problem Formulation}
\label{sec:problem}

This work introduces and addresses the novel problem of \textit{Open Intent Discovery} in written, asynchronous text conversations. The objective of this problem is to accurately determine the existence of one or more \textit{actionable intents} from the text input, and subsequently identify and extract all the possible user intents. 
These intents may be underlying goals, activities or tasks that a user wants to perform or have performed.
We therefore define an intent as a text phrase consisting of two parts: (i) an \textit{action}, which is a word or phrase representing a tangible purpose, task or activity which is to be requested or performed, and (ii) an \textit{object}, which represents those entity words or phrases that the action is going to act or operate upon.  
A similar definition has been used in prior literature to define intention posts in social media and discussion forums~\cite{wang2015mining,chen2013identifying}.
For instance, the intent of the author of the text \textit{``Please make a 10:30 sharp appointment for a haircut"} is to make or schedule a haircut appointment. The intent consists of an action \textit{``make"} and an object \textit{``appointment"}, \textit{``appointment for haircut"}, or \textit{``haircut appointment"}.
Similarly, consider the utterance \textit{``I would like to reserve a seat and request a special meal on my flight"}. In this case, the actions are \textit{``reserve"} and \textit{``request"} and the objects are \textit{seat} and \textit{special meal}, for the respective intents of seat booking and meal request.

\begin{figure}[t]
\centering
  \includegraphics[width=\linewidth,height=4.5cm,keepaspectratio]{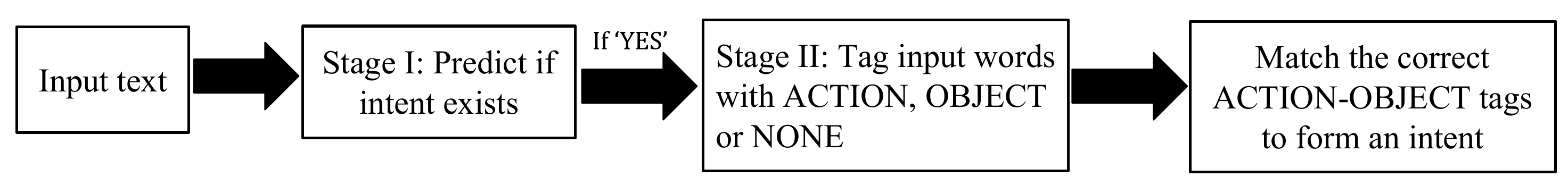}
    \caption{Overview of our \mname{} intent discovery approach}
    \label{fig:pipeline}
\end{figure}

Following our two-part definition of an intent, we formulate the problem of open intent discovery as a sequence tagging problem. We consider three tags: \textsc{Action}, \textsc{Object}, and \textsc{None} (denotes all the remaining words or phrases in the text utterance that are neither an \textsc{Action} nor an \textsc{Object}). A user intent then consists of a matching pair of an \textsc{Action} phrase and an \textsc{Object} phrase. 
Employing such a consistent, generic representation enables our \mname{} framework to identify and extract all possible forms of user intentions which fit in this format. These include previously unseen intent types that were not encountered while training, unlike a classifier that can only address a pre-defined set of intent categories.
This can also help us discover \textit{multiple} possible user intents from a single text inputs and not just a single intent, unlike most of the current literature.
This is crucial since user queries or utterances can often consist of more than one intent in them such as a single main or principal intent, and a few more interlinked, accompanying intents. There could also be multiple tasks that may need to be accomplished together (e.g. \textit{reserve seat} and \textit{request special meal}).  
Besides, having a common format to represent an intent contributes immensely in finding user intents irrespective of their target domain or topic.

We observe from the previous illustrations that the \textsc{Action} component of an intent is likely to consist of a verb or infinitive phrase that follows a noun or a subject phrase. Further, the \textsc{Object} component of an intent often comprises of a noun or compound noun (i.e., an expression with multiple nouns) phrase, possibly qualified by adverbs or adjectives.
Nevertheless, we cannot simply use a part-of-speech (POS) tagger or a language dependency parser for the purpose of identifying intents due to the following reasons. 
First, a POS tagger or a parser cannot distinguish between the various \textsc{Action}-\textsc{Object} pairs present. They will identify all pairs, including irrelevant ones that are merely part of the descriptive text and are not associated with user intent. 
They will hence suffer from a low precision problem, as we empirically show in Table~\ref{tab:step2}.
Second, corresponding \textsc{Action} and \textsc{Object} phrases may be spatially distant from each other in the input text and may even span multiple sentences (see Table~\ref{tab:attention}). 
Having said that, we do notice the efficacy of initially pre-training the weights of the model of the second stage of \mname{} with the verb-object tags obtained from a dependency parser, as we show empirically in Table~\ref{tab:step2}. It helps our model learn generic indicators for various kinds of intents, independent of the topic or domain of the input text. We then fine-tune our model with labeled data specific to our problem. Pre-training can also help overcome the issue of overfitting in sequence learning models, especially if there is insufficient annotated training data.

\mname{} (Figure~\ref{fig:pipeline}) works in two stages for input utterances: 

\begin{enumerate}
\item \textbf{Open Intent Existence Recognition}: The first stage uses a two-layered Bi-LSTM to semantically encode the input utterances. It then employs a binary sigmoid classifier to predict whether an actionable user intent exists within the text input. If not, our method ends here. Otherwise, it moves to the second stage of \textit{extracting} the intents from the text.

\item \textbf{Open Intent Extraction}: The second stage builds upon state-of-the-art sequence tagging systems that utilize bidirectional RNNs and CRFs~\cite{huang2015,lample2016,ma2016end}. It uses a CRF on top of a Bi-LSTM accompanied by adversarial training and attention, to extract actionable user intents from the text input.

\end{enumerate}

In the subsequent sections~\ref{sec:intentexistence} and~\ref{sec:intentextraction}, we describe both these stages.

\section{\mname{} Stage I: Open Intent Existence}
\label{sec:intentexistence}

\begin{figure}[t]
\centering
  \includegraphics[width=\linewidth,height=5.5cm,keepaspectratio]{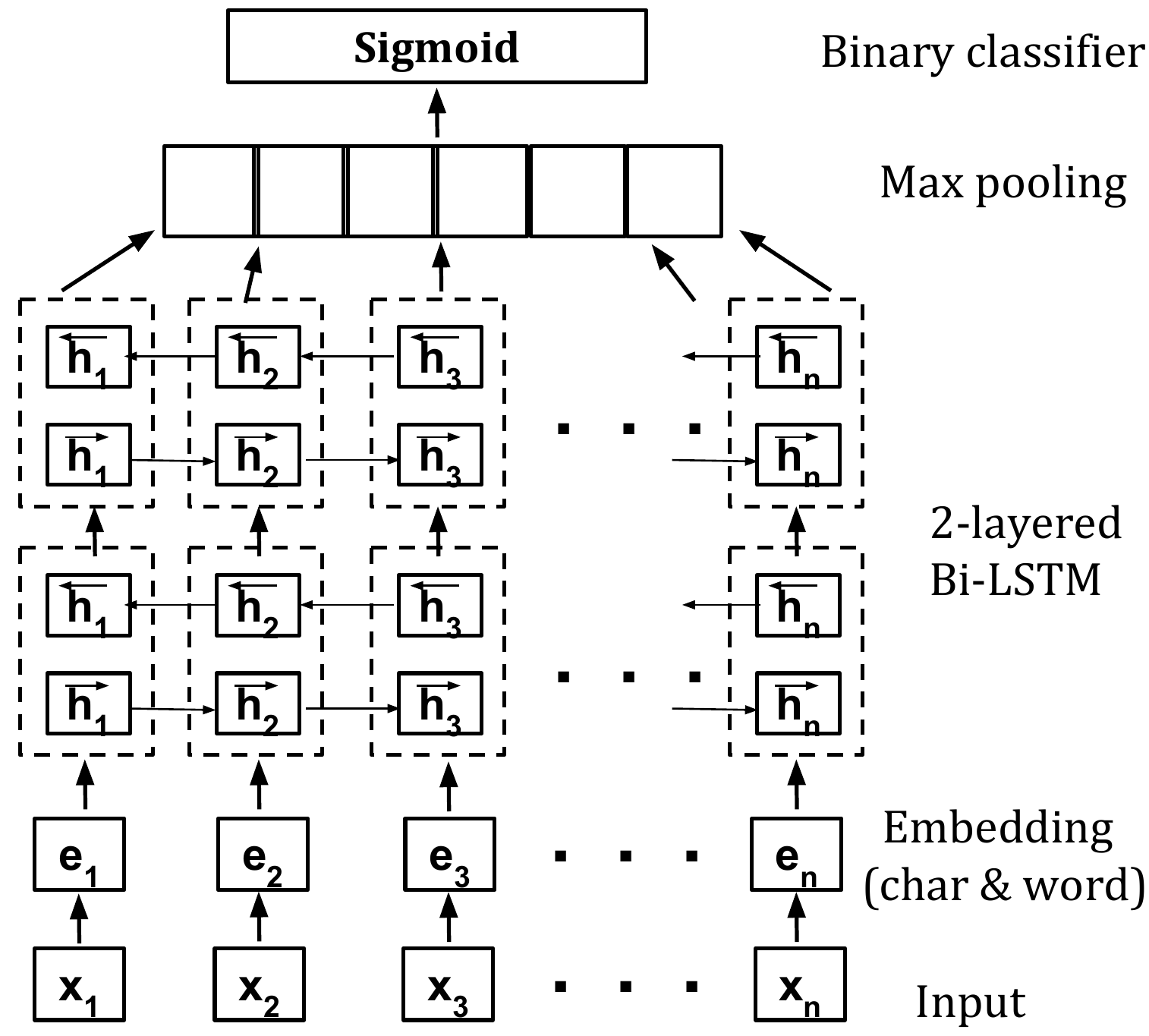}
    \caption{\mname{} Stage I model: open intent existence}
    \label{fig:intentclassifier}
\end{figure}

Given an input text $x$ consisting of a sequence of words $[x_1, x_2, ..., x_n]$, we first transform it into a feature sequence by constructing the character level representation of each word $x_i$. The reason for this is that incorporating character level representations of words via convolutional neural networks (CNNs) can boost the effectiveness of sentence representations, by capturing morphological information present in the language~\cite{zhang2015,ma2016end}.
For this purpose, we build a CNN consisting of convolutional and max pooling layers with dropout~\cite{srivastava2014dropout}, similar to Huang et al~\cite{huang2015}. 
We also obtain word level GloVe embeddings~\cite{pennington2014glove} for each token from a pre-trained model that has been trained on Common Crawl, a giant corpus of web crawled data. Such low-dimensional and dense embeddings are highly effective in capturing both syntactic and semantic information. 
Nevertheless, character-level information can often be overshadowed by word-level embeddings if both are simply concatenated to produce a combined representation for each word. Hence, we adopt a \textit{highway network}~\cite{srivastava2015highway} to retain the impact of the both kinds of embeddings, and merge them in a balanced manner. 

Let $e^{cw}_i$ be the concatenation of the character and word level representations $e^c_i$ and $e^w_i$ of the word $x_i$. The combined embedding $e$ from the highway network is then given by:

\begin{center}
$e = r \odot \tanh (W_H e^{cw} + b_H) + (1-r) \odot e^{cw} $ \\ \smallskip
$r = \sigma(W_R e^{cw} + b_R)$
\end{center}
where $\tanh$ is the hyperbolic tangent function, $\odot$ denotes the element-wise multiplication operation, $W_R$ and $W_H$ are weight matrices, and $b_R$ and $b_H$ are bias vectors. $r$ (transform gate) and $1-r$ (carry gate) are non-linear transformation functions indicating the proportion of output produced by transforming the input and carrying it.
Every word $x_i$ is thus transformed into a $d$-dimensional embedding $e_i$. This embedding layer forms the first layer of our model architecture, and serves as input for the next layer, namely a Bi-LSTM layer.

Bidirectional LSTM~\cite{graves2013} based models have been used for various sequence modeling domains where it is often beneficial to utilize both the past and future context. They are effective in encoding sentences by considering the order of their constituent words and can account for long-range word dependencies.
Figure~\ref{fig:intentclassifier} displays the architecture of the first stage of \mname{}. The Bi-LSTM layer contains two sub-layers for the forward ($\overrightarrow{h_t}$) and backward ($\overleftarrow{h_t}$) sequence contexts respectively. $\overrightarrow{h_t}$ and $\overleftarrow{h_t}$ are both generated based on the recurrences of an LSTM cell~\cite{hochreiter1997}, and their concatenation gives the combined output $h_t$ at time step $t$.
We use a stack of two Bi-LSTM layers to generate a sequence of word-level representations $[h_1, h_2, ..., h_n]$. 
To obtain a fixed length vector from the Bi-LSTM output, we perform one-dimensional max pooling over all $h_t$'s, followed by a sigmoid output layer. It performs a binary prediction of whether an actionable intent is present in the text input or not. 
We use 400 hidden LSTM units with L2 regularization. We apply dropout to the Bi-LSTM and max pooling layers with a probability value of 0.5, to avoid overfitting and co-adaptation of the hidden units. We train the model via the Adam~\cite{kingma2014adam} optimizer with gradient clipping to optimize the binary cross entropy loss function. We set the initial learning rate to 0.001 with a decay of 0.05.
If the first stage of our \mname{} framework predicts the input utterance as containing an intent, it moves to the second stage, i.e. extracting the actual actionable intents from the input utterance.

\section{\mname{} Stage II: Open Intent Extraction}
\label{sec:intentextraction}

\begin{figure}[t]
\centering
  \includegraphics[width=\linewidth,height=5cm,keepaspectratio]{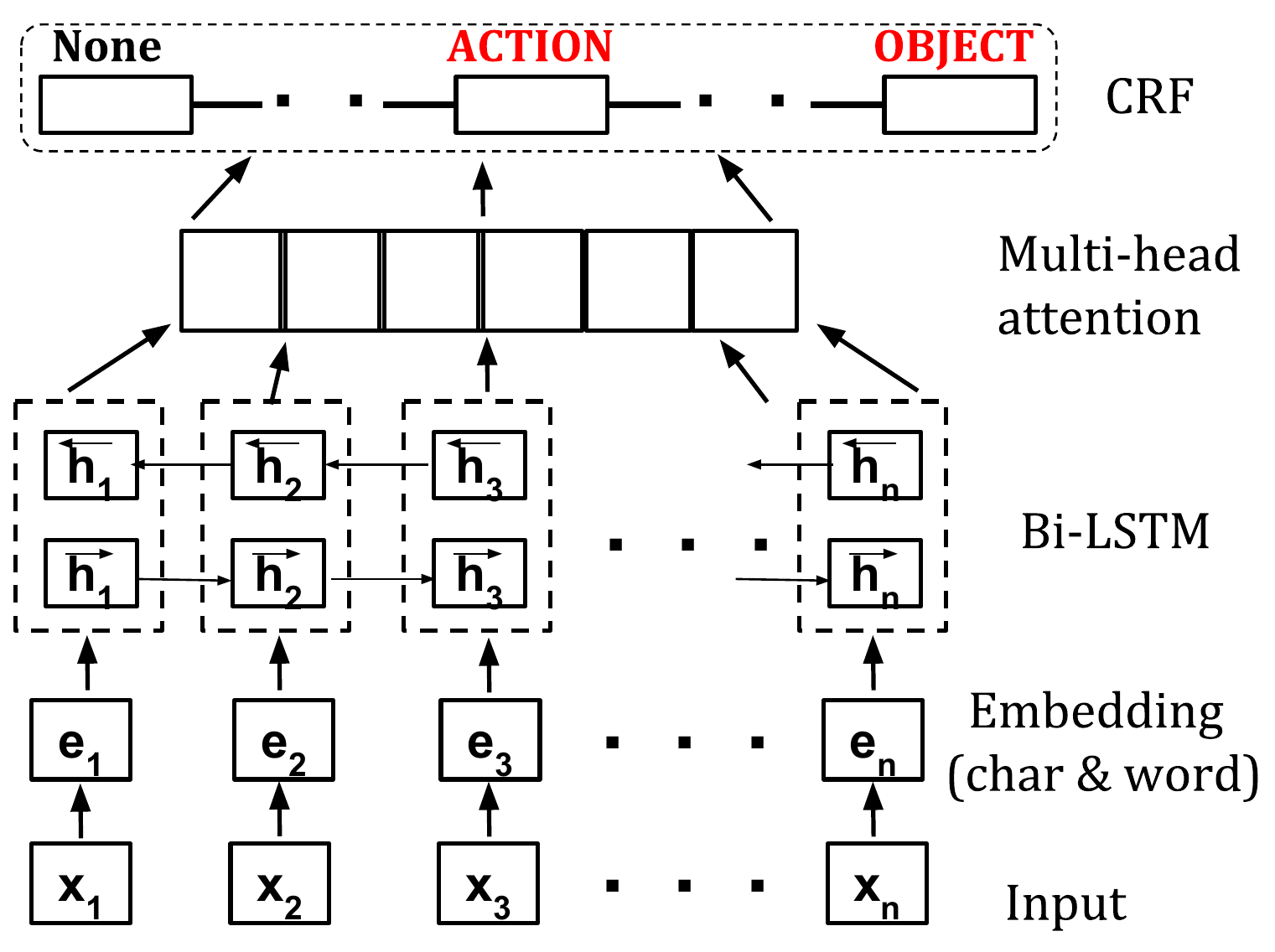}
    \caption{\mname{} Stage II model: open intent extraction}
    \label{fig:intentextractor}
\end{figure}

This is the second stage of our \mname{} intent discovery framework. The initial layers of this intent extraction model (Figure~\ref{fig:intentextractor}), namely the input and embedding layers are constructed similar to stage I (Section~\ref{sec:intentexistence}). The Bi-LSTM structure on top of the embedding layer is also the same, except for the number of its layers (two Bi-LSTM layers for stage I, and one Bi-LSTM layer for stage II). The remaining components of the open intent extraction model are detailed below.

\subsection{Adversarial Training}
\label{sec:adversarial}
Adversarial training~\cite{goodfellow2015,miyato2016adversarial} is a useful technique to regularize predictive models and improve their robustness to small input perturbations. It has also been applied in the literature for the purpose of domain adaptation, i.e., to discover features and structures that are common across multiple domains~\cite{ganin2014unsupervised,kim2017adversarial,liu2017multi}. Since both the above are goals of this work, we utilize adversarial training to enhance our model. We generate \textit{adversarial} input examples that are very close to the original inputs and should yield the same labels, yet are likely to be mispredicted by the current model. These examples are created by adding small \textit{worst case} perturbations or noise to the inputs in the direction that significantly increases the model's loss function. Our model is then trained on the mixture of original and adversarial examples to improve its stability to input perturbations.
Since adversarial training considers continuous perturbations to inputs, we add adversarial noise at the embedding layer, as in~\cite{miyato2016adversarial}. 

Let an input text $x = [x_1, \dots, x_n]$ be represented by an embedding $e$, as detailed in Section~\ref{sec:intentexistence}. We generate its worst case perturbation $\eta$ of a small bounded norm $\epsilon$, which is a hyperparameter to be tuned. It maximizes the loss function $\mathcal{L}$ of the current model with parameters $\theta$ as follows:

\begin{center}

$\eta_{adv} = \argmax \limits_{||\eta|| \leq \epsilon} \mathcal{L}(e+\eta; \theta) $ \\ 

\end{center}

Since the exact computation of $\eta_{adv}$ is intractable in complex neural networks, we use the first order approximation via the fast gradient method~\cite{goodfellow2015,miyato2016adversarial} to obtain an approximate worst case perturbation of norm $\epsilon$. We also normalize the word and character embeddings, so that the model does not trivially learn the embeddings of large norms and make the perturbations insignificant~\cite{miyato2016adversarial}.

\begin{center}
    $\eta_{adv} = \epsilon \frac{g}{||g||}$; where $g = \nabla_e(\mathcal{L}(e; \theta))$  \\   \smallskip

$e_{adv} = e + \eta$  \\ \smallskip
$\mathcal{L}' = \alpha \mathcal{L}(e; \theta') + (1-\alpha) \mathcal{L}(e_{adv}; \theta')$
\end{center}

\noindent where $e_{adv}$ represents an adversarial example generated from embedding $e$. $\mathcal{L}(e; \theta')$ and $\mathcal{L}(e_{adv}; \theta')$ represent the loss functions from the original training instance and its adversarial transformation respectively. $\alpha$ is a weighting parameter. 
The new loss function $\mathcal{L'}$ can be optimized in the same way as the original loss $\mathcal{L}$.

\subsection{Attention Mechanism}

We employ attention to select and focus on the important and essential hidden states of the Bi-LSTM layer. In particular, we use a multi-head self-attention mechanism~\cite{vaswani2017attention,im2017distance,tan2017deep,lin2017structured} that jointly attends to information at different positions of the input sequence, with multiple individual attention functions and separately normalized parameters called \textit{attention heads}. This enables it to capture different contexts in a fine-grained manner and learn long-range dependencies effectively.
Each attention head computes a sequence~$z$ from the output~$h=[h_1, h_2, ..., h_n]$ of the Bi-LSTM layer by projecting it to a key~$k$, a value~$v$, and a query~$q$ via distinct affine transformations with ReLU activations~\cite{glorot2011}. 
The attention weights~$a_{ijp}$ for attention head~$p$ between word tokens $i$ and $j$ are computed as:

\begin{center}
$a_{ijp} = \text{softmax}(\frac{q_{ip}^T k_{jp}}{\sqrt(d)})$ \\ \smallskip

$z_{ip} = \sum_j v_{jp} \odot a_{ijp}$ \\ \smallskip
$z_i = \oplus z_{ip}; \forall p$
\end{center}

\noindent Here $\odot$ denotes an element-wise product and \textit{softmax} indicates the softmax function along the $j$-th dimension. The individual attention head outputs $z_{ip}$ are concatenated into $z_i$ for token $i$.
The scaled dot product above enhances the optimization process by better distributing the gradients and flattening the softmax function~\cite{vaswani2017attention}. 

\subsection{Sequence Tagging via CRFs}

The output of the attention layer serves as input to the next layer of our \mname{} intent extraction model, namely a CRF~\cite{lafferty2001}. CRFs effectively utilize the correlations between labels in a sequence neighborhood to predict the best label sequence for a given input. 
As mentioned earlier, the task of the CRF layer is to predict one of three tags for each word of the input sequence: \textsc{Action}, \textsc{Object}, or \textsc{None}.
The input to the CRF layer is the sequence $z = [z_1, z_2, ..., z_n]$ from the attention layer. 
$y$ represents a certain output label sequence for $z$, and $Y'(z)$ represents the possible set of label sequences. The conditional probability function for the CRF, $P(y|z; W, b)$, over all possible label sequences $y$ given input sequence $z$ is given by:

\begin{center}
$P(y|z; W, b) = \frac{\prod\limits_{i=1}^n \Psi_i (y_{i-1}, y_i, z)}{\sum \limits_{y' \in Y'(z)} \prod\limits_{i=1}^n \Psi_i (y_{i-1}', y_i', z)}$  
\end{center}

\noindent where $\Psi_i (y_{i-1}', y_i', z) = exp(W_{y',y}^T z_i + b_{y',y})$ are potential functions to be learned. $W_{y',y}^T$ and $b_{y',y}$ are weight and bias matrices corresponding to the label pair $(y', y)$.
In this work, we use linear chain CRFs trained via maximum conditional log-likelihood estimation. 

\subsubsection{Enhancing CRFs via Additional Constraints}
\label{sec:constraint}

The Viterbi algorithm~\cite{forney1973viterbi} used for decoding the CRF layer only considers interactions between sequentially adjacent tag labels. However, we encounter additional restrictions or constraints in our problem. 
First, we want to ensure that the CRF never predicts only an \textsc{Action} tag or an \textsc{Object} tag, since our definition mandates the occurrence of both an action and the object it acts upon to constitute a valid intent. 
Next, it is often useful to identify \textit{intent indicator} phrases that suggest the presence of an intent in the corresponding text, or are characteristic of an action following them. Since it is challenging to construct a comprehensive list of all such intent indicators, we pick a small number of highly indicative phrases~\cite{gupta2014identifying,wang2015mining} (see Section~\ref{sec:stage1results} for examples). For each such phrase, we selectively choose candidates having labelled intent tags in a small contextual neighbourhood following the intent indicator. In this work, we use up to five words as the neighborhood length. 

These constraints operate at the level of the fully inferred sequence, and cannot be easily integrated into the Viterbi decoding algorithm by straightforward techniques like modifying its transition matrix~\cite{kristjansson2004interactive}. We circumvent this in two ways during the tag inference phrase of the CRF. 
The first is to use a \textit{beam search} based approach, which penalizes the sequences in the beam not satisfying the aforementioned constraints, and falls back to using the next most probable tag predictions. 
In the second approach, we adopt the idea of reducing the Viterbi decoding algorithm to a graph shortest path problem, solvable by Integer Linear Programming~\cite{roth2005integer}. The authors map a sequence of length $n$ with $m$ possible tag labels to a graph with $nm+2$ nodes and $(n-1)m^2 + 2m$ edges. We extend this formulation by expressing our constraints as linear inequalities, and use this as the decoding algorithm for our CRF layer.
The performance of these enhancements is shown in Table~\ref{tab:step2}.

\subsection{Generating Intents from Tag Sequences}
\label{sec:combinetag}
Once the CRF predicts \textsc{Action}, \textsc{Object} and \textsc{None} tags for each word in the input text, our final step is to pair up and match the corresponding \textsc{Action} and \textsc{Object} tag phrases to generate coherent and meaningful user intents. As specified earlier, we define an intent as a combination of \textsc{Action} tagged phrases followed by \textsc{Object} tagged phrases. We develop two techniques for this purpose.
First, we employ the simple but highly effective technique of linking \textsc{Action} and \textsc{Object} tagged phrases with respect to their word-based proximity in the input text. We assume that related action-object phrases are likely to occur spatially close to each other in this distance-based heuristic. For instance, in our earlier example, the action `\textit{reserve}' is more likely to match with the object \textit{seat}', than with the object `\textit{special meal}'. 
However, this assumption may not always hold true, depending on the way the text has been worded. 

This brings us to our second technique of matching appropriate \textsc{Action}-\textsc{Object} tagged phrases, by learning a multi-layer perceptron (MLP) classifier. The input features for the MLP consist of the sum of the pre-trained GloVe embeddings (alluded to in Section~\ref{sec:intentexistence}) of the words in the potential \textsc{Action}-\textsc{Object} intent phrase, concatenated with the normalized value of the spatial word distance between the \textsc{Action} and \textsc{Object} phrases in the original input text. These features account for both the word proximity of the constituent intent terms, and their semantic likelihood of co-occurring in a single phrase.
The input to the MLP during both training and testing is thus the feature representation of all possible paired up combinations of the predicted \textsc{Action} and \textsc{Object} tagged phrases.
On top of the input layer, the MLP contains two fully connected hidden layers of rectified linear units, followed by a fully connected layer of size one. This outputs a score $y_{mlp}$ for each potential \textsc{Action}-\textsc{Object} pair under consideration, indicating the likelihood of combining them to produce an intent. 
We train the MLP with a margin-based hinge loss function $\mathcal{L}_{hinge}$, that maximizes the separation margin between the true and the highest scoring incorrect \textsc{Object} phrase option for the current \textsc{Action} phrase.

\begin{center}
\smallskip
$\mathcal{L}_{hinge} = \sum_{i=1}^{N} max(0, 1-y_{mlp}^{(i)} \ \cdot \ y_{mlp}'^{(i)}) $
\end{center}
where 
$y_{mlp}^{i} = \pm 1$ is the intended output of the MLP classifier, indicating whether the $i$-th \textsc{Action}-\textsc{Object} pair is a correct match or not, and $y_{mlp}'^{(i)}$ is the raw output value of its decision function.
We present the performance of both the above techniques in Table~\ref{tab:step2}.

Our self-attention based Bi-LSTM-CRF open intent extraction model thus makes use of semantic information from the previous and future time steps, and dependency constraints learned and enforced by the CRF; to predict intents for an input text utterance.
Multi-head self-attention enables it to learn dependencies between distant words, possibly across sentences, effectively.
Adversarial training acts as a powerful regularizer for our model, contributing to its robustness and resilience to user intents from diverse domains.

\section{Evaluation}
\label{sec:evaluation}

\subsection{Data Collection}

We collected about 75K questions with their top correct answer, on various topical categories and task domains, posed by users on the question-answer forum \url{www.stackexchange.com}. We then formulated an Amazon Mechanical Turk crowd sourcing experiment to annotate 25K of these questions with up to three intents (specifically, action phrases and their corresponding object phrases), that the crowd workers felt were most important or relevant.

\begin{table}[!t]
\caption{Statistics of our curated Stack Exchange dataset}
\label{tab:stats}
\small
\begin{tabularx}{\linewidth}{L{0.82}|L{0.33}|L{0.44}|L{0.4}}
\hline
\textbf{Name of Genre} & \textbf{No. of utterances} & \textbf{Avg length of utterance} & \textbf{Vocab size per genre} \\
\hline
Data science & 8184 & 60 & 11561 \\
Software engineering & 7114 & 60 & 23417\\
Web apps & 7048 & 50 & 28906\\
Webmasters & 7524 & 56 & 18688\\
Sharepoint & 9366 & 60 & 40094\\
Productivity & 8968 & 60 & 9529\\
Development ops & 1660 & 60 & 1871\\
Open data & 2166 & 60 & 7952 \\
Server fault & 7772 & 53 & 16047\\
Life hacks & 1836 & 50 & 7837 \\
DIY & 2378 & 35 & 4140\\
CRM software (name omitted for blind review) & 11723 & 60 & 47219\\
\hline
\end{tabularx}
\end{table}

\begin{table}[!t]
\caption{Results for open intent existence prediction}
\label{tab:step1}
\small
\begin{tabularx}{\linewidth}{L{1.65}|L{0.35}}
\hline
\textbf{Approach} & \textbf{F1-score} \\
\hline
Random forest classifier (features from Section~\ref{sec:stage1results}) & 0.66 \\
MLP classifier (features from Section~\ref{sec:stage1results}) & 0.74 \\
CNN classifier & 0.83 \\
GRU classifier & 0.813 \\
LSTM classifier & 0.857 \\
Bi-LSTM classifier & 0.864 \\
\mname{} (stage I) classifier & \textbf{0.91}  \\
\hline
\end{tabularx}
\end{table}

For the first stage of \mname{}, i.e. recognizing the existence of an intent in an utterance, we generate two classes from our labeled intent data to provide as training data to the binary classifier. As the positive class (i.e. containing an intent), we use the questions from Stack Exchange that were labeled with user intents by the crowd workers. 
We assume for the purposes of this work that only the questions asked by individuals contain an actionable intent or task that a user desires or requests to be performed. The answers to these questions largely contain informative descriptions, reasons for specific events and occurrences, or methods to carry out activities. The authors of such answers often refer to task actions that need to be performed by \textit{others}, while we seek to identify the personal intentions of users themselves.
Hence, we use these answers as instances of the negative class, i.e. text inputs not containing any intent.
This method of selecting the training and test data contributes to the robustness and efficacy of our binary classifier by minimizing the semantic gap in the topics constituting the two classes. It also ensures that our classifier correctly learns semantic, domain-independent indicators of the presence of an intent.

For the second stage of our \mname{} framework, we use the remaining 50K unlabeled questions for the purpose of unsupervised pre-training.
We first generate \textit{verb} and \textit{object} semantic parsing tags for these texts via the Stanford CoreNLP dependency parser~\cite{manning2014stanford}. We employ the words tagged as verbs and objects as proxies for the \textsc{Action} and \textsc{Object} tagged phrases that compose an intent.
We then fine-tune our model by further training it with the intent-labeled data, tagged with \textsc{Action} and \textsc{Object} phrases by our annotators. 

In Table~\ref{tab:stats}, we depict the different genres of our annotated Stack Exchange dataset. The last row shows a commercial Customer Relationship Management (CRM) software, whose name we omit for the purposes of the double-blind review.
For each genre, we show the number of input question-answer texts collected, the average length of a question text, and the genre's vocabulary size. Each question text serves as input to the second stage of our framework. There are hundreds of unique intent types per category.
We choose Stack Exchange as our data source because of its long and verbose questions with background details and multiple intent actions scattered throughout the text. This is unlike the commonly used SNIPS~\cite{coucke2018} and ATIS~\cite{hemphill1990,dahl1994} datasets for intent detection. These contain relatively shorter and crisper text on pre-defined intent categories, such as booking travels and weather inquiries. Having said that, we do analyze \mname{}'s performance on the SNIPS dataset in Section~\ref{sec:casestudy}.


\subsection{Results}

\subsubsection{\mname{} Stage I: Open Intent Existence Prediction}
\label{sec:stage1results}

Table~\ref{tab:step1} depicts the performance of the first stage of \mname{}, i.e., discovering the existence of an intent. Two simple yet effective baselines we compared against include a random forest and an MLP classifier with two hidden layers of 100 ReLU units each. We framed their input features to capture various lexical and syntactic properties of natural language~\cite{qadir2011}. 
These include: 
(i) the number of nouns and verbs in the text, 
(ii) does the utterance end in a noun or adjective, 
(iii) does the utterance begin with a verb or modal (e.g. \textit{will, would, could} etc) word, 
(iv) the count of `Wh-' markers (e.g. \textit{who, what, how} etc) and question marks that signify a question, 
(v) does the text utterance contain personal pronouns, 
(vi) is there a first-person pronoun (e.g. \textit{i, we}) within a three-word window of an infinitive verb phrase (`to' followed by a verb) in the utterance; and 
(vii) are there phrases indicating an `action plan', or intent (e.g.\ \textit{plan to, want to,} or \textit{would like to}) -- also used in Section~\ref{sec:constraint}.

Additional baselines include sigmoid classifiers deriving features from a CNN with 2D convolution and pooling operations~\cite{kim2014convolutional}, and RNNs (a GRU~\cite{tang2015document}, LSTM and a Bi-LSTM) with a single hidden layer, max pooling and dropout. The first stage of \mname{} achieves an F1-score above 90\%, outperforming the baselines by at least 5\%.

\begin{table*}[!t]
\caption{Performance of various approaches on open intent extraction (stage II). P, R, F1 denote precision, recall and F1 score.} 
\label{tab:step2}
\small
\begin{tabularx}{\linewidth}{L{0.9}|L{0.26}|L{0.26}|L{0.25}|L{0.33}}
\hline
\textbf{Approach} & \textbf{ACTION P/R/F1} & \textbf{OBJECT P/R/F1} & \textbf{Intent P/R/F1} & \textbf{Semantic similarity}  \\
\hline
Phrases following intent indicators (from Section~\ref{sec:constraint}) & 0.65/0.59/0.62  & 0.6/0.54/0.57  & 0.63/0.56/0.59 & 0.67 \\
Stanford CoreNLP dependency parser (SC) & 0.56/0.49/0.52 & 0.51/0.43/0.47 & 0.53/0.45/0.49  & 0.59 \\
\hline
\hline
\mname{} (att+adv) + train on SC + w-dist & 0.60/0.53/0.56 & 0.57/0.47/0.52 & 0.58/0.49/0.53 & 0.70 \\
\mname{} (att+adv) + train on mturk + w-dist & 0.75/0.59/0.66  & 0.74/0.52/0.61 & 0.74/0.55/0.63  & 0.74 \\
\mname{} (att) + pretrain on SC + fine tune on mturk + w-dist & 0.78/0.62/0.69 & 0.79/0.56/0.66 & 0.78/0.58/0.67 & 0.80 \\
\mname{} (adv) + pretrain + fine tune + w-dist & 0.81/0.60/0.68 & 0.76/0.54/0.63 & 0.78/0.56/0.65 & 0.77 \\
\mname{} (att+adv) + pretrain + fine tune + w-dist & 0.84/0.66/0.73 & 0.81/0.63/0.71 & 0.82/0.64/0.72 & 0.83 \\
\mname{} (att+adv+beam-CRF) + pretrain + fine tune + w-dist & 0.84/0.70/0.76 & 0.81/0.67/0.73 & 0.82/0.68/0.74  & 0.84 \\
\mname{} (att+adv+constr-CRF) + pretrain + fine tune + w-dist & 0.84/0.72/0.77 & 0.81/0.67/0.73 & 0.82/0.69/0.75 & 0.85 \\
\hline
\mname{} (att+adv) + pretrain + fine tune + MLP & 0.84/0.68/0.75 & 0.81/0.67/0.73 & 0.82/0.67/0.74  & 0.84 \\
\mname{} (att+adv+beam-CRF) + pretrain + fine tune + MLP & 0.84/0.72/0.77 & \textbf{0.81/0.69/0.75} & \textbf{0.82/0.70/0.76}   & \textbf{0.86} \\
\mname{} (att+adv+constr-CRF) + pretrain + fine tune + MLP & \textbf{0.84/0.73/0.78} & 0.81/0.68/0.74 & \textbf{0.82/0.70/0.76} & \textbf{0.86} \\

\hline
\end{tabularx}
\end{table*}

\subsubsection{\mname{} Stage II: Open Intent Extraction}
\label{sec:stage2results}

\noindent \textbf{Baselines:} Table~\ref{tab:step2} presents a comparison between various approaches for the task of open intent extraction, i.e stage II of our framework. 
In the first approach (first row), we simply return as intents the five-word phrases following the occurrence of any intent-indicator phrases (described in Sections~\ref{sec:constraint} and~\ref{sec:stage1results}). The second baseline is the verb-object tuples learned by the Stanford CoreNLP dependency parser~\cite{manning2014stanford}, which we use as proxies for \textsc{Action} and \textsc{Object} tagged phrases respectively.
The subsequent rows comprise different variants of \mname{}. `\textit{att}' and `\textit{adv}' indicate the presence of attention and adversarial training respectively (from Section~\ref{sec:intentextraction}). `\textit{train on SC}' denotes model training only on verb-object tags obtained from the dependency parser. `\textit{train on mturk}' indicates model training only on the crowd worker annotated intent data.
Except for the third and fourth rows of Table~\ref{tab:step2}, all other variants of \mname{} are first pre-trained on the verb-object tags learned by the dependency parser, followed by fine-tuning on the intent annotated utterances.
`\textit{beam-CRF}' and `\textit{constr-CRF}' refer to the two CRF enhancing strategies from Section~\ref{sec:constraint} of (i) considering a beam of probable tag sequences, and (ii) incorporating additional constraints into the CRF decoding algorithm.
\textit{`w-dist'} and `\textit{MLP}' refer to the two techniques from Section~\ref{sec:combinetag} of matching appropriate \textsc{Action}-\textsc{Object} phrases to create a holistic intent based on (i) word proximity in the input text, and (ii) the score learned by the MLP classifier. 

We reiterate that \mname{} cannot be directly compared with existing intent detection techniques (e.g.~\cite{hakkani2016multi,liu2016attention,zhang2018joint,xia2018}) since these are formulated as classification problems, and classify intents into limited, pre-defined categories. They also require sufficient labeled training data for most (if not all) categories.
Contrarily, \mname{} handles thousands of distinct intent classes, and has no restrictions on the number of training examples needed per unique intent type. 

\smallskip
\noindent \textbf{Performance Metrics}: We use the precision, recall, F1-score and semantic similarity metrics to evaluate the approaches in Table~\ref{tab:step2}.
The second column only considers the prediction performance of the \textsc{Action} tags for each word of the input utterance, whereas the third column only assesses the \textsc{Object} tags.
The fourth column displays the results considering the combination of both tag types to create an intent. 
The last column of semantic similarity computes the average of the cosine similarities between the embeddings of the predicted and actual (ground truth) intents. The embedding for each predicted and true intent phrase is acquired by averaging the pre-trained GloVe embeddings~\cite{pennington2014glove} of their constituent words. 
We ignore the words whose embeddings do not exist.

We observe a significant improvement of \mname{} of over 15\% in terms of F1-score and semantic similarity, compared to the simple intent-indicator based model and the Stanford parser (first two rows of Table~\ref{tab:step2}). Utilizing the dependency parser data as a pre-training step for the weights of our model, followed by continuing the training on the actual intent-labeled data improves the F1-score by at least 6\%.
Enhancing the CRF decoding algorithm with added constraints (\textit{beam-CRF} and \textit{constr-CRF}) benefits the F1-score further by 2-5\%.
We find a performance difference of $\leq$ 3\% between using the word proximity heuristic (\textit{w-dist}), and the MLP classifier for matching \textsc{Action} and \textsc{Object} phrases. 
Overall, \mname{} trained with attention, adversarial training and CRF enhancements outperforms all baselines in Table~\ref{tab:step2}, with an intent F1 score of 76\%, and a semantic similarity of 86\% between the true and predicted intents.

\subsubsection{Capacity of Domain Adaptation}

\begin{table}[!t]
\caption{Studying \mname{}'s domain adaptation capability. The training set for each test domain row includes labeled data from all domains in Table~\ref{tab:stats}, except its own. `+' shows results on including some test domain data while training.}
\label{tab:domain}
\small
\begin{tabularx}{\linewidth}{L{1.05}|L{0.35}|L{0.391}|L{0.105}|L{0.104}}
\hline
\textbf{Test Domain Name} & \textbf{Intent F1} & \textbf{Intent F1+} & \textbf{Sim} & \textbf{Sim+} \\
\hline
Data science & 0.76 & 0.8 & 0.84 & 0.88 \\
Software engineering & 0.69 & 0.74 & 0.81 & 0.86 \\
Web apps & 0.73 & 0.77 & 0.83 & 0.88\\
Webmasters & 0.75 & 0.79 & 0.83 & 0.86\\
Sharepoint & 0.71 & 0.76 & 0.82 & 0.85\\
Productivity & 0.73 & 0.78 & 0.81 & 0.86\\
Development ops & 0.71 & 0.73 & 0.78 & 0.83\\
Open data & 0.69 & 0.73 & 0.84 & 0.87\\
Server fault & 0.67 & 0.72 & 0.75 & 0.8\\
Life hacks & 0.635 & 0.7 & 0.74 & 0.8\\
DIY & 0.72 & 0.76 & 0.81 & 0.86 \\
CRM software (name omitted)  & 0.79 & 0.83 & 0.88 & 0.91 \\
\hline
\end{tabularx}
\end{table}

Encountering newly emergent niche domains or genres for which little to no labeled intent utterances are available is a fairly common real-world scenario. However, it is a time-consuming and labor-intensive process to obtain sufficient domain-specific annotated data for the purpose of model training and development. 
It is thus desirable to adapt and generalize an existing trained model with minimum re-training effort, each time a new domain with potentially new intents is added. 
We investigate the capability of our \mname{} open intent discovery framework in adapting and transferring knowledge across distinct conversational domains or subjects.

In Table~\ref{tab:domain}, we consider several different test domains. We train \mname{} on utterances from the remaining domains other than the test domain. The second and fourth columns assess our method's F1-score and semantic similarity, in predicting the correct intents per unique domain. The definitions of these metrics are the same as in Section~\ref{sec:stage2results}.
The third and fifth columns of Table~\ref{tab:domain} indicate the respective F1-score and semantic similarity achievable for the given test domain, when \mname{} is trained using labeled data from the testing domain as well. 
The difference in both metrics with and without using training data from the test domain is $\leq 5\%$, for most domain topics. Only the \textit{Life Hacks} domain suffers a loss of 6.5\% in terms of F1-score when we eliminate the data from this domain while training our model. 
Interestingly for the domain of \textit{DIY}, its training data is dominated by other semantically distinct domains, as can be seen in Table~\ref{tab:domain}. However, \mname{} still attains a good F1-score of 72\%, only 4\% lesser than what was possible if \textit{DIY} domain data was included in the training set.
These results show that \mname{} can easily and effectively be generalized to low-resource domains with minimal manual effort, to detect actionable intents in newly emerging domains with potentially novel intents.

\subsubsection{Effect of Human-Annotated Training Data Size}

\begin{figure}[t]
\centering
  \includegraphics[width=\linewidth,height=3.5cm,keepaspectratio]{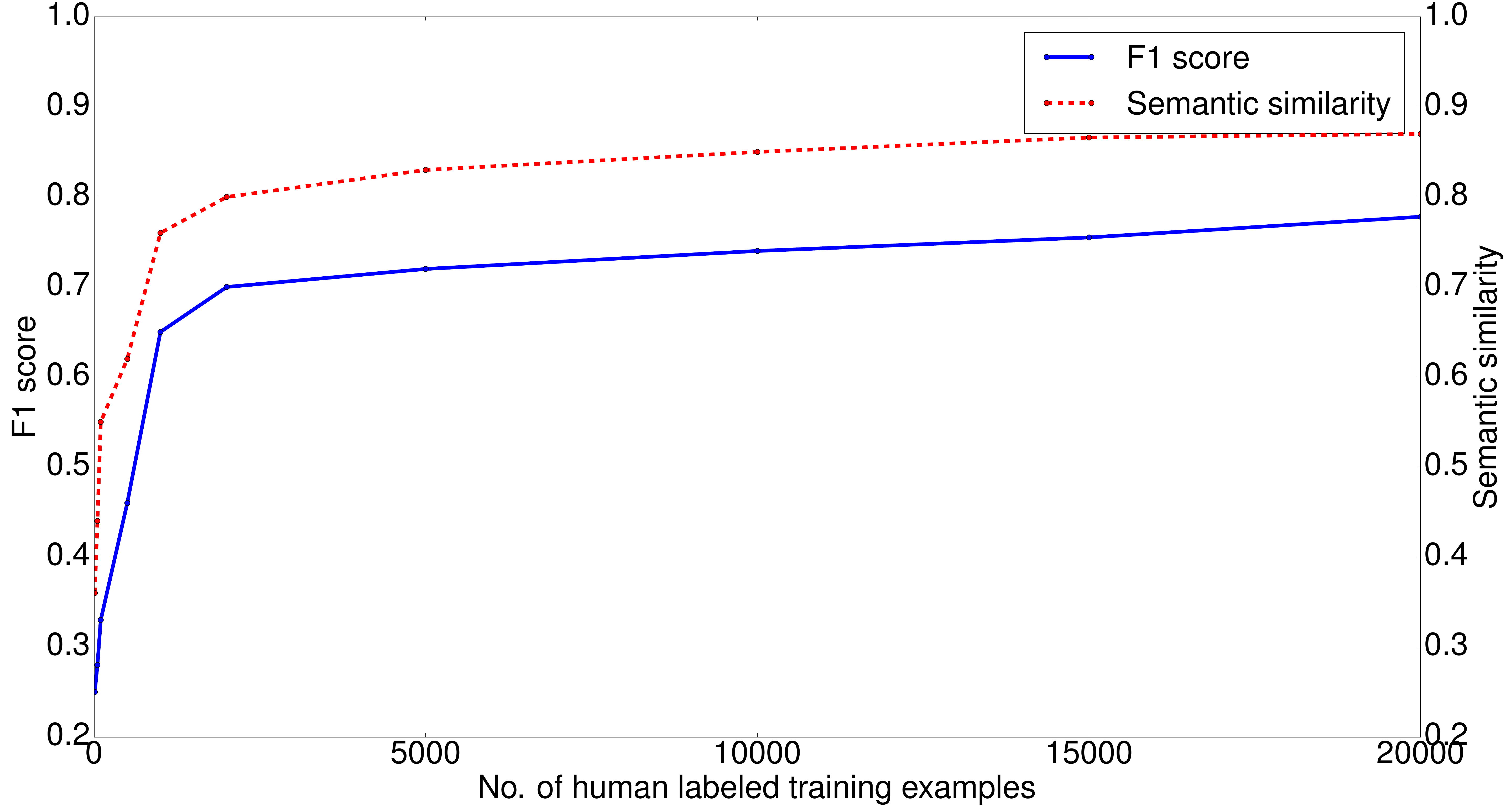}
    \caption{Effect of varying the amount of human labeled data}
    \label{fig:numhumanlabel}
\end{figure}

In the third and fourth rows of Table~\ref{tab:step2}, we showed that training our \mname{} model with absolutely no human-labeled intent data is detrimental to its performance.
We next examine in more detail, the effect of using varying amounts of human annotated intent data while training our model. Note that we also use the tags from the dependency parser to pre-train our model.
Figure~\ref{fig:numhumanlabel} shows the F1-score and semantic similarity values for the predicted intents achieved by \mname{}, as the number of human annotated training instances varies. Both metrics are represented on the dual y-axes, by the blue and red plots respectively. We find that both plots are monotonically increasing. When the number of human labeled training instances across various domains is less than $1000$, both metrics are below 50\%. The F1-score and semantic similarity rise to about 70\% and 75\% respectively at $1000$ annotated training examples. Since we test \mname{} on our annotated dataset that has $12$ distinct domain topics, this translates to slightly more than $80$ labeled examples per unique domain. Beyond this point, there is a steady performance improvement, with a less sharper gain than earlier.
These observations reinforce \mname{}'s domain adaptation capability and show that it does not require a large number of human labeled examples to attain a high performance.

\subsubsection{Role of Attention}

\begin{table}[!t]
\caption{Effect of attention on utterances. Darker highlight shows higher attention. Boldface denotes presence of intent.}
\label{tab:attention}
\small
\begin{tabularx}{\linewidth}{L{1.53}|L{0.47}}
\hline
\textbf{Input Text Utterance} & \textbf{Intents} \\
\hline
Is it \hlc[pink3]{possible to \textbf{navigate}}\hlc[pink2]{back} in <XXX>\textbf{\hlc[pink3]{to}\hlc[pink1]{previous page}} after\hlc[pink2]{save}\hlc[pink3]{processing?} ... I have a page where I\hlc[pink0]{click on} a link and use navigateURL ... \hlc[pink4]{want to} \hlc[pink3]{be able to}\hlc[pink4]{go back}\hlc[pink4]{to} the\hlc[pink2]{previous calling page} and \hlc[pink1]{\textbf{complete} the \textbf{processing} of} the\hlc[pink2]{\textbf{save}}... & navigate previous page, complete processing save \\
\hline
The "Your tweets retweeted" page on Twitter... \hlc[pink2]{\textbf{find} out all} the \textbf{users} who\hlc[pink4]{\textbf{retweeted}} a tweet of mine? ... how many people have\hlc[pink3]{retweeted a}\hlc[pink2]{tweet} and\hlc[pink2]{what their}\hlc[pink1]{\textbf{Twitter IDs} are}?  & find retweeted Twitter IDs \\
\hline
Is there a WordPress plugin that will\hlc[pink4]{\textbf{tweet when}}\hlc[pink2]{a scheduled} post is posted? I know there are tons ... that will tweet\hlc[pink3]{when} you\hlc[pink3]{\textbf{publish} a post}, but none\hlc[pink0]{I have tried} will do it on a\hlc[pink3]{\textbf{scheduled post}}.  & tweet when publish scheduled post \\
\hline
\hlc[pink4]{How can I}\hlc[pink2]{\textbf{keep} my}\hlc[pink3]{\textbf{phone}}\hlc[pink2]{\textbf{from}} just\hlc[pink1]{\textbf{falling}}\hlc[pink1]{over}when watching videos? ... also\hlc[pink3]{want to \textbf{have}}\hlc[pink1]{my}\hlc[pink2]{\textbf{hands}}\hlc[pink3]{\textbf{free} to do}other things ... but I find due to most\hlc[pink0]{phones not being} particularly `grippy' it is hard to lean them up ... any nifty\hlc[pink0]{life hacks} for this? & keep phone from falling, have hands free \\
\hline
I'm starting a micro-school...\hlc[pink3]{I want}\hlc[pink4]{to}\hlc[pink2]{\textbf{manage sick notes}}\hlc[pink3]{and}\hlc[pink2]{\textbf{absences}} ...\hlc[pink3]{How can I}\hlc[pink2]{\textbf{synchronize} one}\hlc[pink1]{\textbf{central}}\hlc[pink0]{Google \textbf{Calendar}} that only administrators have access... Parents should\hlc[pink0]{be able to schedule} future\hlc[pink2]{absences and} excuse past absences... & manage sick notes, manage absences, synchronize central calendar \\
\hline
\end{tabularx}
\end{table}

Table~\ref{tab:step2} indicates that the presence of attention lends \mname{} an F1 score gain of at least 4\%. We further explore \mname{}'s capability of identifying relevant and meaningful semantic features from its input utterances, which contribute in discovering open intents. We examine and visualize in Table~\ref{tab:attention} the self-attention values for specific utterances from our Stack Exchange dataset.
For the sake of brevity we display truncated versions of the text inputs in the first column, and the second column shows their associated user intents. A darker colored highlight on a specific utterance word indicates that it receives higher attention, and consequently plays a greater role in \mname{}'s decision of intent discovery. Input utterance words that constitute intents are marked in boldface.  
In all cases, we observe that words semantically related to and contributing to at least one user intent are successfully identified by an attention head. For instance, the second row of Table~\ref{tab:attention} demonstrates the significance of \textit{`find out', `retweeted', `tweet'} and \textit{`what their Twitter IDs are'} in deciding the user intent of \textit{``find retweeted Twitter IDs"}. 
The attention heads are attentive to intent indicator phrases that are likely to precede an actionable intention, such as \textit{`possible to', `want to be able to', `how can I'} and \textit{`I want to'}. 
Words that may represent an action or an object but are irrelevant to the user's intent (e.g. \textit{`click on a link'} and \textit{`use navigateURL'} in the first row, \textit{`I find'} and \textit{`lean them up'} in the last row) also do not receive a high attention score. 
Further, our attention mechanism can capture the dependency between distant intent words, such as \textit{`find'} and \textit{`retweeted'} in the second row and \textit{`publish'} and \textit{`scheduled'} in the fourth row. It also associates the action \textit{`manage'} with two objects, \textit{`sick notes'} and \textit{`absences'}, generating the intents \textit{``manage sick notes"} and \textit{``manage absences"}. 

\subsection{Case Studies}
\label{sec:casestudy}

We now test the efficacy of \mname{} on two additional real-world datasets. Note that these serve as test instances for our models that have already been trained on our curated Stack Exchange dataset.

\subsubsection{SNIPS NLU Benchmark Dataset}


\begin{figure}
\begin{subfigure}{.245\textwidth}
  \centering
  \includegraphics[width=\linewidth]{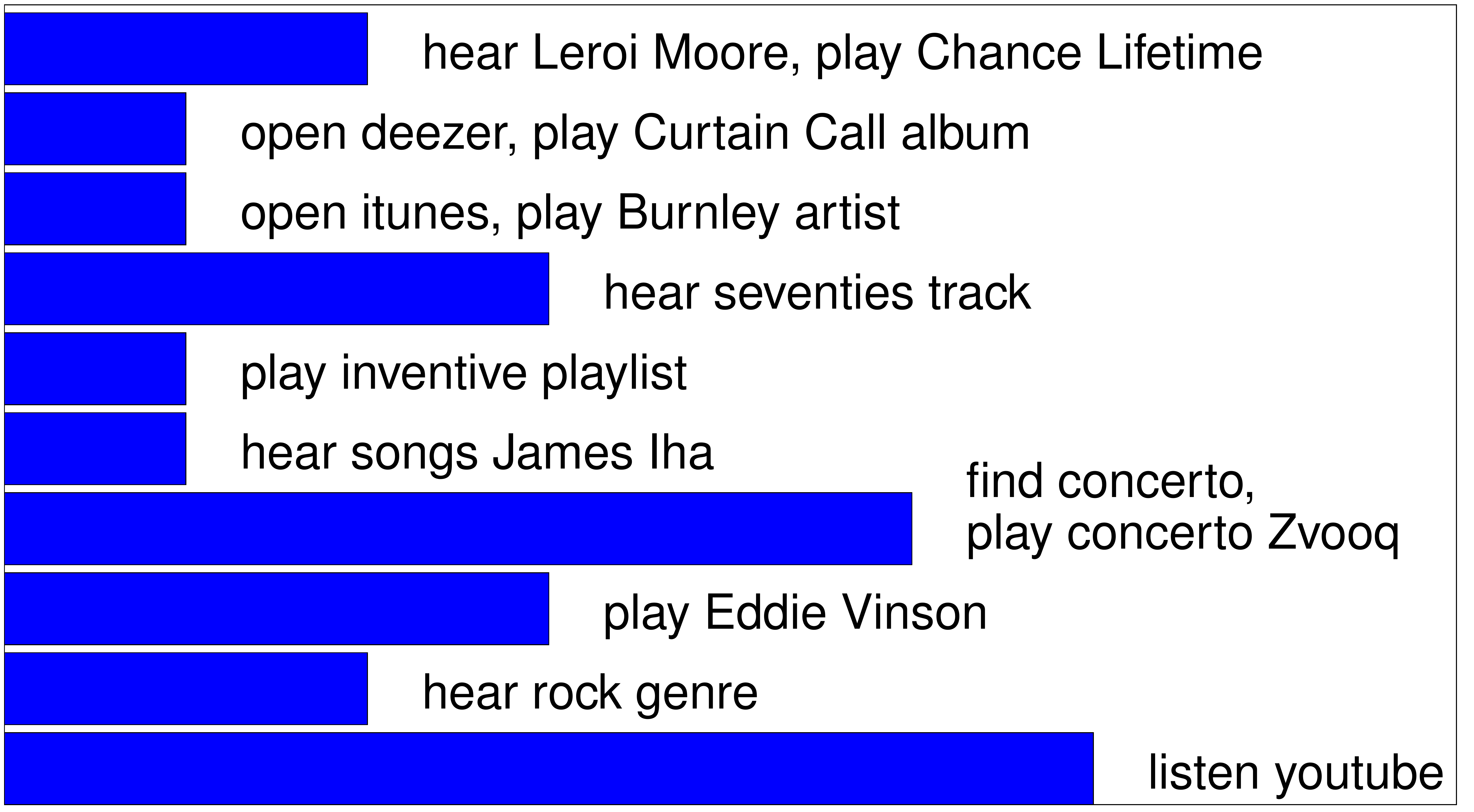}
  \caption{SNIPS: PlayMusic}
\end{subfigure}%
\begin{subfigure}{.245\textwidth}
  \centering
  \includegraphics[width=\linewidth]{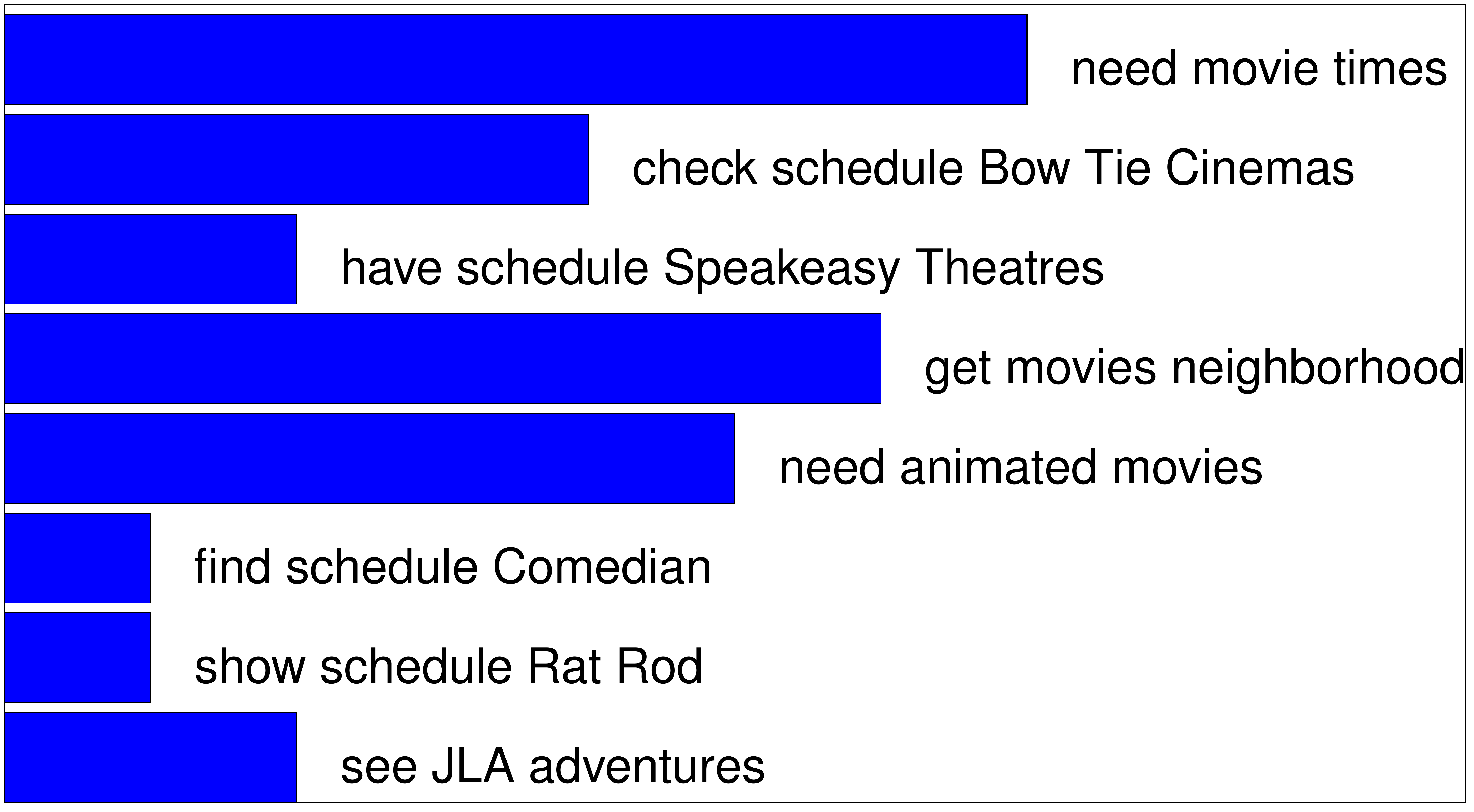}
  \caption{SNIPS: SearchCreativeWork}
\end{subfigure}
\begin{subfigure}{.245\textwidth}
  \centering
  \includegraphics[width=\linewidth]{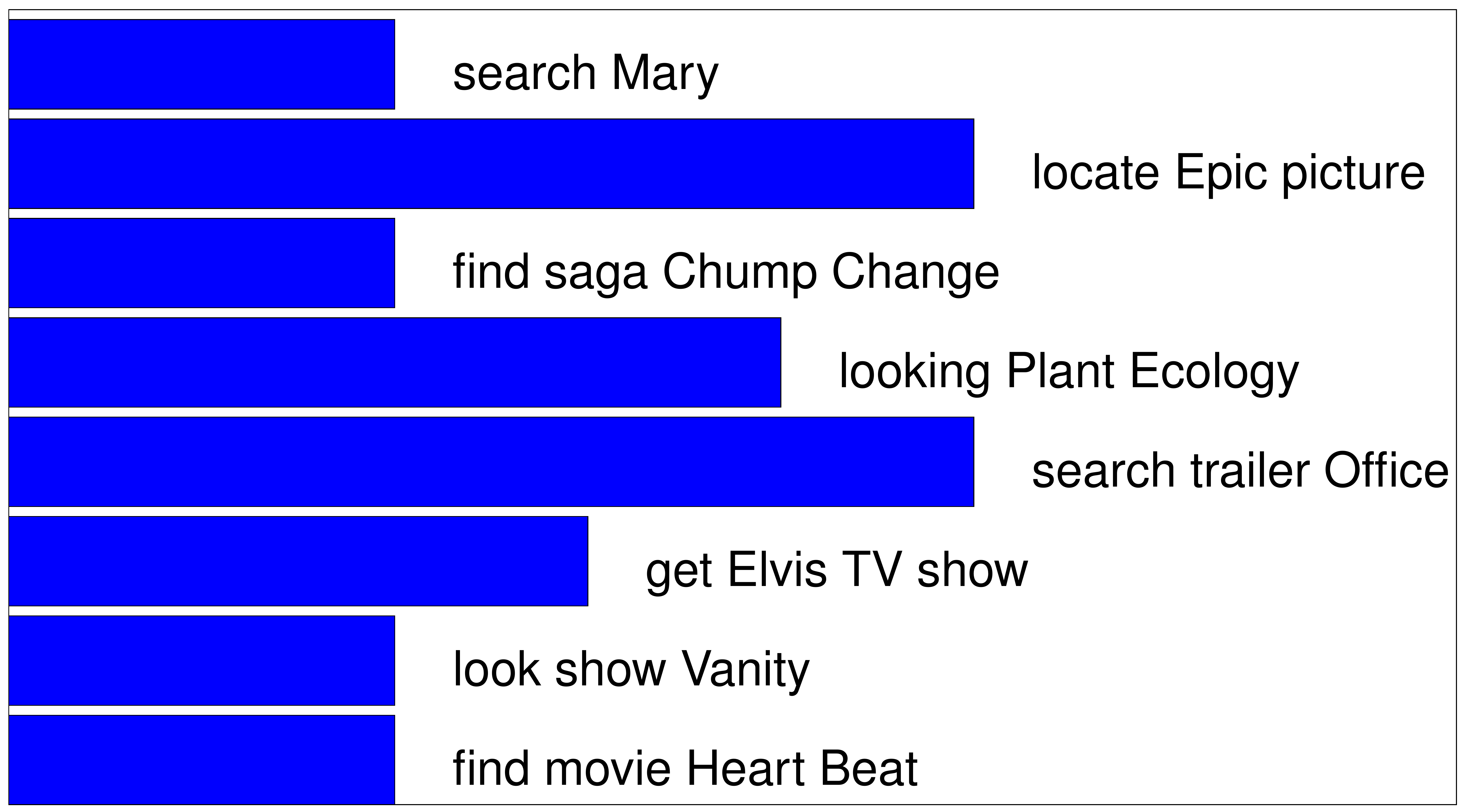}
  \caption{SNIPS: SearchScreeningEvent}
\end{subfigure}%
\begin{subfigure}{.245\textwidth}
  \centering
  \includegraphics[width=\linewidth]{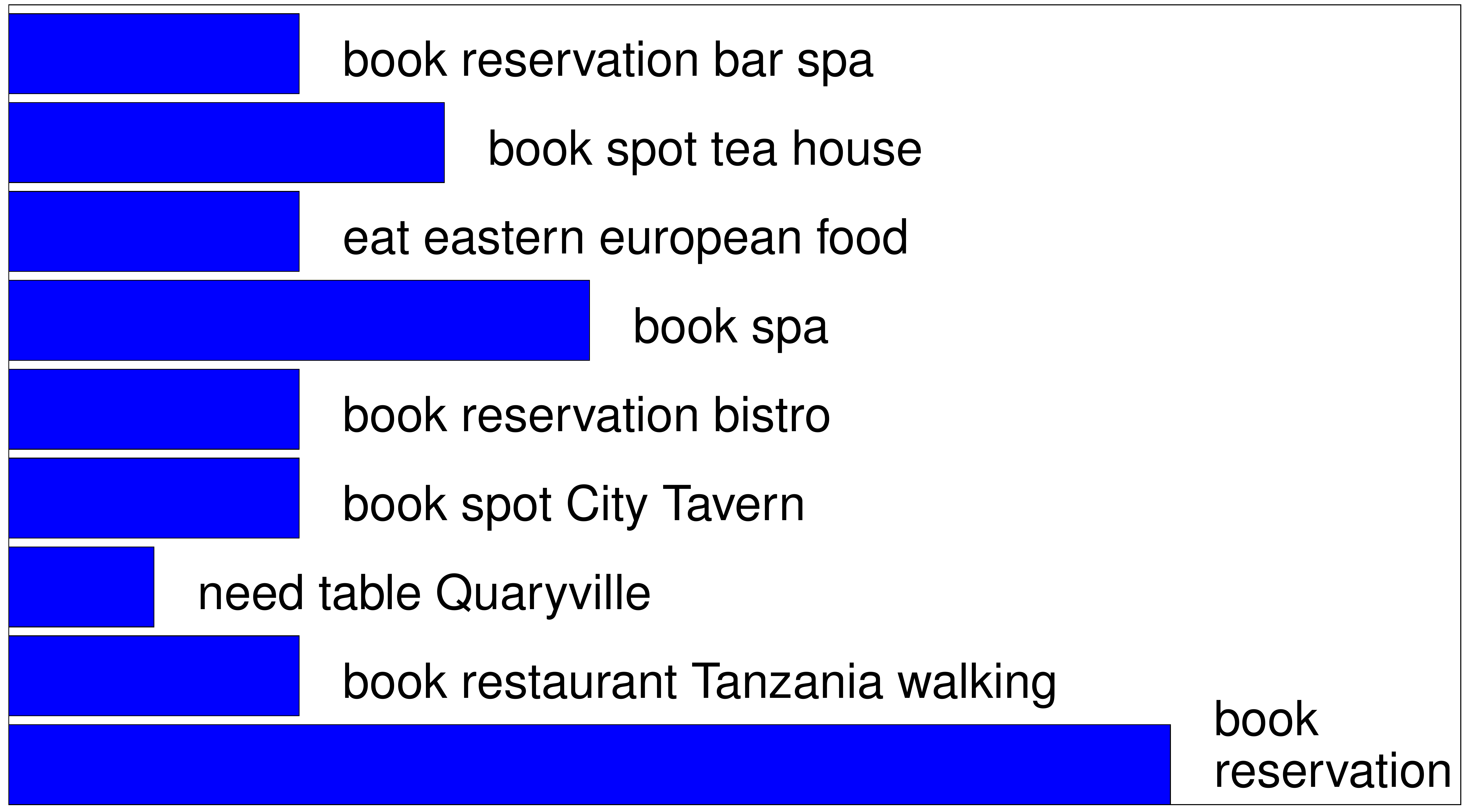}
  \caption{SNIPS: BookReservation}
\end{subfigure}
\caption{Fine-grained intents discovered by \mname{} for four high-level intent categories in the SNIPS NLU dataset}
\label{fig:casestudy}
\end{figure}

\begin{table}[!t]
\caption{Performance of \mname{} on a technical support dialog snippet. Words that make up intents are shown in bold.}
\label{tab:casestudy}
\small
\begin{tabular}{|>{\arraybackslash \sffamily}p{8.3cm}|}
\hline
\textsc{\textbf{User}}: Can someone please help? I'm trying to \textbf{fix} a \textbf{broken ubuntu}. \\
\textsc{\textbf{Agent}}: ... how did you break it? \\
\textsc{\textbf{User}}: i'm on the cd and i'm trying to \textbf{mount} and then \textbf{chroot} my \textbf{hd}, which worked fine. I installed some new libs and now it no longer reboots. \\
\textsc{\textbf{User}}: what's the easiest way to \textbf{get} a \textbf{working boot} on my drive again? \\
\textsc{\textbf{Agent}}: ... sounds like something might be screwed up in your /etc/apt/sources.list file, if it's failing on apt-get update \\
\textsc{\textbf{User}}: how can i \textbf{fix} my \textbf{sources.list file}? \\
\textsc{\textbf{Agent}}: open /etc/apt/sources.list. see if you notice any obvious errors \\
\textsc{\textbf{User}}: a question on the mounting issue - when i loaded the cd, my local hard drive was mounted in media, can't i just use that as the chroot? \\
\textsc{\textbf{Agent}}: ... assuming your flgrx is hosed, move the x conf file out of the way so that the radeon driver will be used instead ... \\
\textsc{\textbf{User}}: ... what do you suggest for a good backup program for ubuntu? \\
\textsc{\textbf{User}}: ... i installed the latest radeon drivers manually. how do i \textbf{upgrade} to the \textbf{newest kernel} and default \textbf{radeon drivers}? \\
\textsc{\textbf{Agent}}: first you'd uninstall 10.6 fglrx driver. then you'd grab the three 2.6.34 deb packages and then install xorg-edgers repo. run grub-update so it finds the new kernel and done.\\
\textsc{\textbf{User}}: where do i \textbf{get} the \textbf{debs}? and i know how to \textbf{uninstall} the \textbf{fglrx drivers} ..., and then do i \textbf{copy} back the \textbf{xorg.conf.original} to xorg.conf? \\
\textsc{\textbf{User}}: ... do i need to add a source to my source list? \\
\textsc{\textbf{Agent}}: yes you need xorg-edgers (google it) \\
\textsc{\textbf{User}}: ok cool. how do i \textbf{get rid} of \textbf{xorg}, or is that already done? \\
\textsc{\textbf{Agent}}: ... if you used jockey-gtk to install fglrx and no other method, then you should be able to use the same method to remove them \\
\hline
\end{tabular}
\end{table}

This is a collection of over 16K crowdsourced queries on seven different topics provided by the commercial company SNIPS, and is widely used to benchmark the performance of automated dialog response agents~\cite{coucke2018}. This dataset has a vocabulary size nearly $20$ times smaller than our curated Stack Exchange dataset. The length of its average input is at least $6$ times shorter. Further, the SNIPS data is both linguistically and semantically less diverse as well as less complex due to the specificity of its constituent topics.  
We test \mname{} trained on completely unrelated domains from Table~\ref{tab:stats} on the SNIPS dataset. We show in Figure~\ref{fig:casestudy} sample intents it discovers for the four distinct intent categories of \textit{PlayMusic, SearchCreativeWork, SearchScreeningEvent} and \textit{BookRestaurant}. The length of the bars represents the relative frequency of that particular intent in the input data.
We observe that our method can be highly beneficial in drilling down further into the high-level intent categories, and understanding and summarizing the exact and specific fine-grained user actionable intents that they comprise of. For instance, in Figure~\ref{fig:casestudy}(a), \mname{} not only identifies the basic intents of `hear song' or `play album' in the \textit{PlayMusic} category; but also tells us that users are interested in singers such as \textit{Leroi Moore, Eddie Vinson} and \textit{James Iha}, song albums like \textit{Curtain Call} or \textit{concerto}, and music platforms like \textit{Youtube} and \textit{Zvooq}.  
Further, though the focus of \mname{} is open intent discovery and not slot filling, in most cases \mname{} can automatically identify important and meaningful accompanying information apart from the user's principal intent. For example, in Figure~\ref{fig:casestudy}(c), a user's overall desire is to search for screenings of a particular event. Our method accurately predicts this via keywords like \textit{search, locate, find} and \textit{look}. Moreover, \mname{} also provides added information on the specific events that need to be searched, such as the \textit{Chump Change} saga and the movie \textit{Heart Beat}.

\subsubsection{Ubuntu Dialog Corpus Chat Logs}

Table~\ref{tab:casestudy} shows a real-world, multi-turn conversation between a user with technical issues (called \textsc{User}), and another who helps resolve them (called \textsc{Agent}). It belongs to the Ubuntu Dialog Corpus~\cite{lowe2015ubuntu}. This dataset contains about one million technical support conversations related to the Ubuntu Linux operating system, and highly resembles real-world dialog exchanges with commercial customer care agents. The original dialog from which this snippet has been truncated contains more than $100$ turns.
In general, such data is asynchronous with several dialog turns. It has diverse and informal user intents, dialog domains and semantic slots; which increases the difficulty of the open intent discovery task. 

Note that \mname{} has been trained on primarily unrelated genres from our Stack Exchange dataset, before testing on this dialog. Our goal here is to understand the intents of the user requesting support (\textsc{User}), and not the one providing it (\textsc{Agent}). The words constituting intents inferred by \mname{} have been highlighted in boldface. 
We observe that though our framework was trained on labeled data which had up to three annotated intents, it is capable of recognizing more than three intents where applicable. \mname{} recognizes the following user intents in the whole conversation: \textit{fix broken ubuntu, mount hd, chroot hd, get working boot, fix sources.list file, upgrade newest kernel, upgrade radeon drivers, get debs, uninstall fglrx drivers, copy xorg.conf original} and \textit{get rid xorg}. 
Using a classification-based intent detection approach, it would be quite difficult to meaningfully categorize such a conversation into a single intent type or category.
Our \mname{} framework on the other hand provides a realistic, fine-grained summary of the action items that a user intends to perform throughout the conversation.
\section{Conclusion}
\label{sec:conclusion}

In this work, we introduced and tackled the problem of open intent discovery. We developed a two-stage novel sequence tagging approach, \textit{\mname{}}, in contrast to the common method of modeling intent detection as a multi-class classification task. Our proposed framework harnesses a Bi-LSTM and a CRF coupled with self-attention and adversarial training. It can extract from user utterances multiple actionable intent types in a consistent format, many of which may be unseen during training. 
We additionally curated a large collection of 25K instances from diverse domains on Stack Exchange, and annotated them for general-purpose intents via crowd sourcing. 
Experiments and case studies on real-world datasets showed substantial improvements of our approach over competitive baselines.
We also demonstrated \mname{}'s ability to generalize and adapt across multiple domains, thereby minimizing the amount of labeled training data for a new task domain.
\mname{} provides an in-depth, fine-grained understanding of users' prospective actions and intentions from their text utterances, which can greatly benefit downstream end-to-end conversational applications. 

Our current work extracts explicitly mentioned user intents from their utterances. Promising future directions could include learning generative models for intents, and inferring implicitly present open intents from users' text or speech. Another interesting direction could be applying \mname{} to informal social media conversations.



\bibliographystyle{ACM-Reference-Format}
\bibliography{final_report}

\end{document}